\documentclass[12pt]{iopart}
\usepackage{graphicx}
\usepackage{color}
\begin{document}

\title{Generation of pulsed bipartite entanglement using four-wave mixing}

\author{Quentin Glorieux$^*$, Jeremy B. Clark, Neil V. Corzo, and Paul D. Lett}

\address{Quantum Measurement Division, National Institute of Standards and Technology\\
and Joint Quantum Institute, NIST and University of Maryland,\\
100 Bureau Dr., Gaithersburg, MD 20899-8424\\
$^*$Contact address : quentin.glorieux@nist.gov}

\begin{abstract}
Using four-wave mixing in a hot atomic vapor, we generate a pair of entangled twin beams in the microsecond pulsed regime near the D1 line of $^{85}$Rb, making it compatible with commonly used quantum memory techniques.
The beams are generated in the bright and vacuum-squeezed regimes, requiring two separate methods of analysis, without and with local oscillators, respectively.
 We report a noise reduction of up to $3.8\pm 0.2$~dB below the standard quantum limit  in the pulsed regime and a level of entanglement that violates an Einstein--Podolsky--Rosen inequality.
\end{abstract}

\maketitle

\section{Introduction}

Non-classical light is a key resource for quantum information and communication \cite{Braunstein:2005wr} as well as precision measurement \cite{Caves:1981vh}.
Just as continuous-wave (CW) squeezing can be useful for noise reduction in gravitational wave detection \cite{Abadie:2011dj,Vahlbruch:2010bp}, a robust source of pulsed squeezed light would be useful in the domain of quantum communication \cite{Braunstein:2005wr}.
Indeed, the ability to store continuous variable entanglement in two separated quantum memories would allow for the entanglement of two macroscopic atomic ensembles \cite{polzik1}, similar to what has been achieved in the single photon regime \cite{Choi:2008wt,Zhao:2009kz}.
An experimental issue is the availability of an entangled light source compatible with a given quantum memory.
The compatibility is often limited due to optical wavelength and bandwidth mismatch between them.

In this paper we focus on generating quantum states of light suitable for Rb atomic memories such as those based on Electromagnetically Induced Transparency or Gradient Echo Memory \cite{Hosseini:2011iv,Hosseini:2011jv,Lvovsky:2009vg} by using a Rb vapor near resonance as a non-linear medium.
Four-wave-mixing (4WM) in Rb vapor has been shown to produce strongly correlated multi-spatial mode "twin beams" \cite{McCormick:2007tl,Boyer:2008ts,Marino:2009wg,Clark:2012wd,Glorieux:2011fj}.
Previous studies of 4WM performed in the nanosecond pulsed regime have demonstrated squeezing \cite{Agha:2010tk,Agha:2011cd}, but the inseparability of such states in the pulsed regime has not yet been established.
Other techniques (e.g. Kerr effect in optical fibers) have successfully reached levels of entanglement that violate an Einstein-Podolsky-Rosen (EPR) inequality \cite{PhysRevLett.86.4267}, and here we use an atomic medium to demonstrate up to $3.8\pm 0.2$~dB of squeezing as well as EPR levels of entanglement between twin beams in the microsecond pulsed regime.
In particular, we describe in detail two methods of analysis useful for characterizing bright beam intensity-difference squeezing and measuring entanglement between vacuum-squeezed twin beams, similar to \cite{Appel:2008vl,Honda:2008cj}.
We do this by taking the Fourier transform of the noise time trace or integrating the photocurrent over the pulse window, respectively.

\section{Four-wave mixing and squeezed light}
\begin{figure}[b]
\centering\includegraphics[width=\columnwidth]{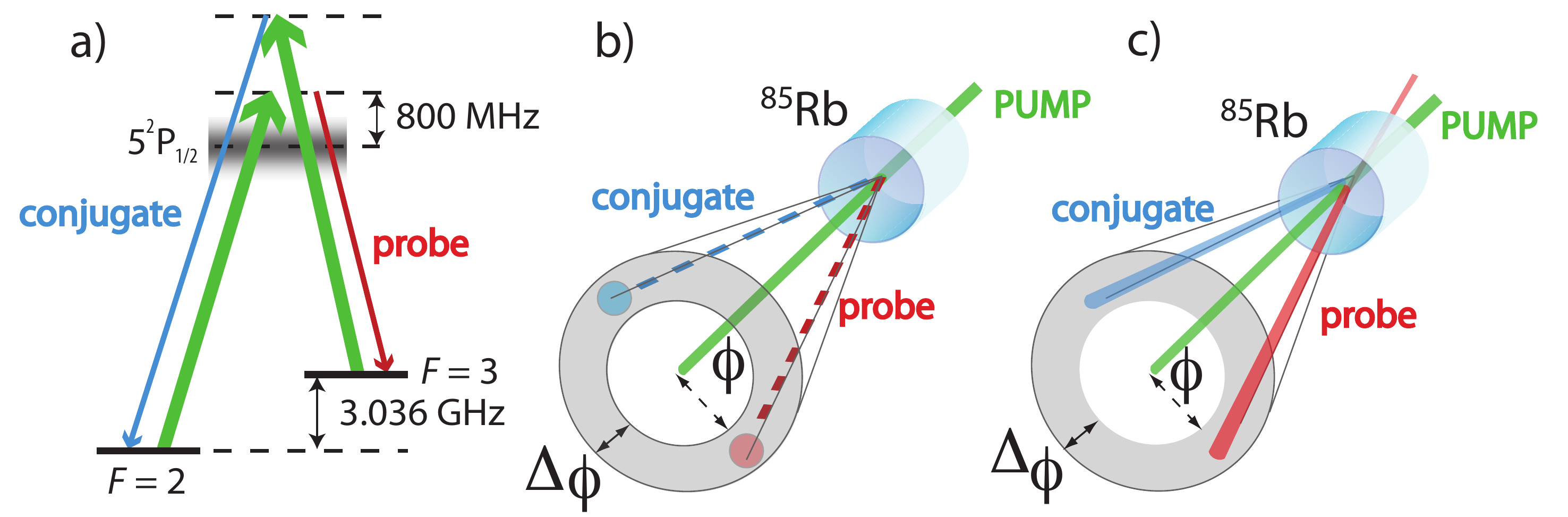}
\caption{\label{fig:4WM}Four-wave mixing in $^{85}$Rb vapor. (a) Energy levels. (b) Vacuum-squeezed twin beam generation.  Probe and conjugate fields are coupled over a range of angles $\Delta \phi$ determined by the phase matching conditions, and selected by the local oscillators used for homodyne detection.  (c) Bright twin beam generation with a coherent state seeding the probe field at $\phi$.}
\end{figure}

Our experimental setup is based on the one described in \cite{Boyer:2008ts}.
We generate strongly correlated CW twin beams using four-wave mixing in a double-$\Lambda$ configuration in a warm $^{85}$Rb vapor.
As illustrated in Fig.~\ref{fig:4WM}, the four-wave mixing (4WM) process converts a pair of photons provided by a strong pump field into a pair of photons emitted into twin fields referred to as the probe and the conjugate.
The pump beam is detuned approximately 800~MHz to the blue of the $\vert5^{2}S_{1/2},F~=~2\rangle \rightarrow \vert5^{2}P_{1/2},F~=~3\rangle$ transition at 795~nm and focused down to a $\frac{1}{e^{2}}$ diameter of 1.2~mm inside a 1.25~cm long $^{85}$Rb cell heated to 110\ensuremath{^\circ}C.
Taken individually, the probe and conjugate beams behave like thermal states exhibiting excess quadrature noise relative to the shot noise limit (SNL).
When compared these twin fields are known to exhibit strong correlations \cite{Glorieux:2010ja}.

To obtain bright twin beams, we stimulate the 4WM process by injecting a coherent state ``seed"  into the probe mode, which produces bright probe and conjugate beams suitable for direct detection with photodiodes (Fig.~\ref{fig:4WM}c).
These bright beams are intensity-difference squeezed, and the squeezing is measured by directly detecting the probe and conjugate intensities with matched photodiodes with quantum efficiencies of 95\%.
The photocurrents are subtracted and the intensity-difference signal is amplified and recorded.

Although easy to implement, direct detection does not provide access to measurements of the phase quadrature, a necessary condition to show entanglement \cite{Hillery:2006je}.
To measure these anti-correlations, a common approach is to perform a simultaneous homodyne detection of each beam  (Fig.~\ref{fig:4WM}b).
The generalized quadrature $\hat{X}_{\theta_i}$ can be defined as $\hat{X}_{\theta_i}=\frac{1}{\sqrt{2}}(\hat{a}^{\dagger}e^{i\theta_i}+\hat{a}e^{-i\theta_i})$, where $i$ can be $p$ or $c$ for the probe and conjugate beams respectively.
The phase of each local oscillator (LO) can select the quadrature to be detected in each beam and accordingly, if the relative phase between the LOs is stable over the detection time, it is sufficient to sweep only one LO phase in order to fully characterize the state.

When the signals obtained by the homodyne detections are subtracted, the noise power of the joint quadrature $\hat{X}^{-}_{\theta}=\hat{X}^{p}_{\theta_p}-\hat{X}^{c}_{\theta_c}$  where $\theta\equiv\theta_{p}+\theta_{c}$ is measured.
In this case it is possible to observe squeezing provided that $\theta\equiv\theta^{-} = 0$.
Similarly, we can measure the sum of the two homodyne signals in order to obtain the joint quadrature $\hat{X}^{+}_{\theta}=\hat{X}^{p}_{\theta_p}+\hat{X}^{c}_{\theta_c}$ which is squeezed if $\theta\equiv\theta^{+}= \pi/2$.
The observation of squeezing on both joint quadratures is sufficient to demonstrate entanglement \cite{Hillery:2006je}.
The degree of entanglement can be quantified by the inseparability $I$, defined as the sum of the noise variances measured for  $\hat{X}^{+}_{\theta^{+}}$ and $\hat{X}^{-}_{\theta^{-}}$ normalized to the SNL : $I=\left<(\Delta X^{-}_{\theta^{-}})^2\right>+\left<(\Delta X^{+}_{\theta^{+}})^2\right>$.
States are said to be entangled if $I<2$ \cite{Duan:2000ts,Simon:2000vh}.

\section{Intensity-difference squeezing}
\begin{figure}[htbp]
\centering\includegraphics[width=0.8\columnwidth]{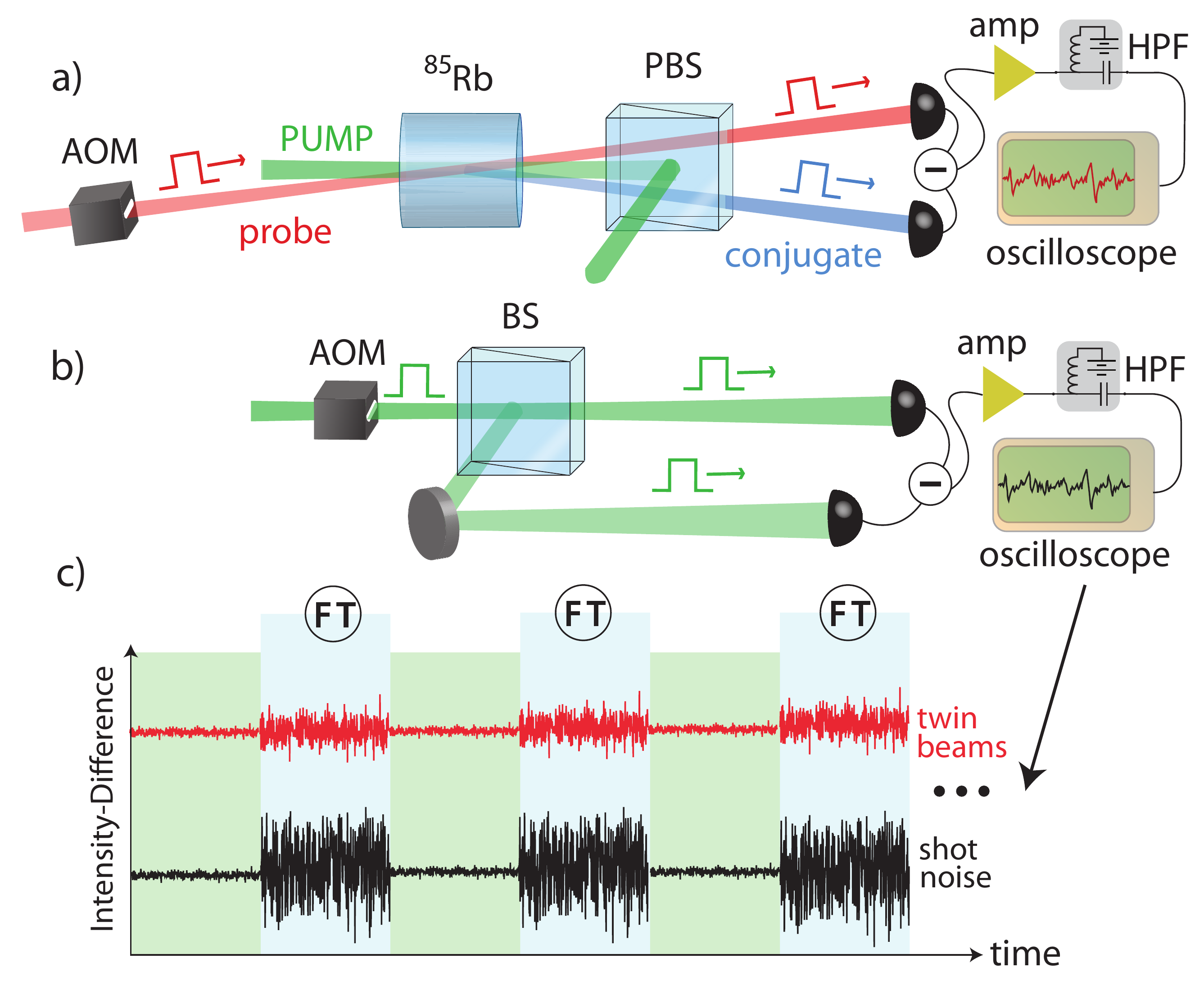}
\caption{\label{fig:setupBB}
Experimental procedure to detect pulsed bright beam intensity-difference squeezing.
PBS, polarizing beamsplitter; BS, 50-50 beamsplitter; HPF, high-pass filter.
(a) Detection of pulsed bright beam intensity-difference noise.
(b) Pulsed shot noise detection.
(c) Analysis procedure to measure intensity-difference squeezing between bright beams.  The pulse train of bright beams yields a subtracted photocurrent signal (red trace) that alternates between the squeezed intensity-difference signal (regions shaded blue) and the electronic noise floor (regions shaded green).  The Fourier transform of the time trace in each blue region is computed and then averaged to obtain the final power spectrum of the intensity-difference noise. Traces in this figure are not experimental results and are included only to clarify the analysis procedure. 
}
\end{figure}

We apply these techniques to the pulsed regime for bright beams and vacuum--squeezed twin beams.
We first measure intensity-difference squeezing between pulsed bright beams as sketched in Fig.~\ref{fig:setupBB}.
A coherent state at the probe frequency is obtained by double-passing an acousto-optic modulator (AOM) coupled to a 1.5 GHz radio frequency (RF) signal.
The RF drive is then pulsed in time, producing a train of pulses for the input seed.
While the applied RF pulses were square pulses, the optical pulses were bandwidth--limited by the 4WM process.
High--frequency components of the pulses are not amplified by the process and have a negligible contribution to the noise measurement.
These pulses are shot noise limited which is an important condition in order to observe intensity--difference squeezing.
The pulse width is 2 $\mu$s and pulses are generated with a period of 10 $\mu$s.
These bright pulses are directly detected with matched photodiodes and the resulting photocurrents are subtracted.
The noise of the subtracted photocurrents therefore alternates between the intensity-difference noise of the squeezed light and the electronic noise floor of the detection system when the pulses are not present (Fig.~\ref{fig:setupBB}).

The arrival of the probe lags the conjugate by approximately 10~ns due to the dispersion in the atomic vapor \cite{vincentPRL}.
This difference in the arrival times generates a voltage spike from the RF amplifier, and the capacitance of the detecting circuit produces a large ringing effect that dominates the intensity difference signal at frequencies below 0.5~MHz.
A delay line does not resolve this issue because the exact time delay between the pulses fluctuates and it is not possible to cancel the voltage spike to the required level.
Although the noise power introduced by the ringing primarily dominates frequencies below 2~MHz, it is large enough to saturate the amplifying circuitry of the detection system.
To suppress this ringing, we included a high-pass filter with a cutoff frequency of 300~kHz in the detection circuit.

We detected the shot noise by pulsing a shot-noise-limited beam picked off from the master laser with the same duty cycle as the seed, splitting the pulse with a 50-50 beamsplitter, and directly detecting with the same detection technique and the same detection circuitry (see Fig.~\ref{fig:setupBB}).
Given the detected intensity-difference noise, the intensity-difference squeezing was computed as indicated in Fig.~\ref{fig:setupBB}c.
Each portion of the amplified photocurrent corresponding to the noise of the twin beam intensity difference (blue regions in Fig.~\ref{fig:setupBB}c) consisted of 200 values sampled over 2 $\mu$s, adequately sampling frequencies over the $\approx$~20~MHz bandwidth of the 4WM process.
We computed the Fourier transform for each blue region in Fig.~\ref{fig:setupBB}c and then averaged $10^3$ power spectra to obtain the final power spectrum.

\begin{figure}[htbp]
\centering\includegraphics[width=0.7\columnwidth]{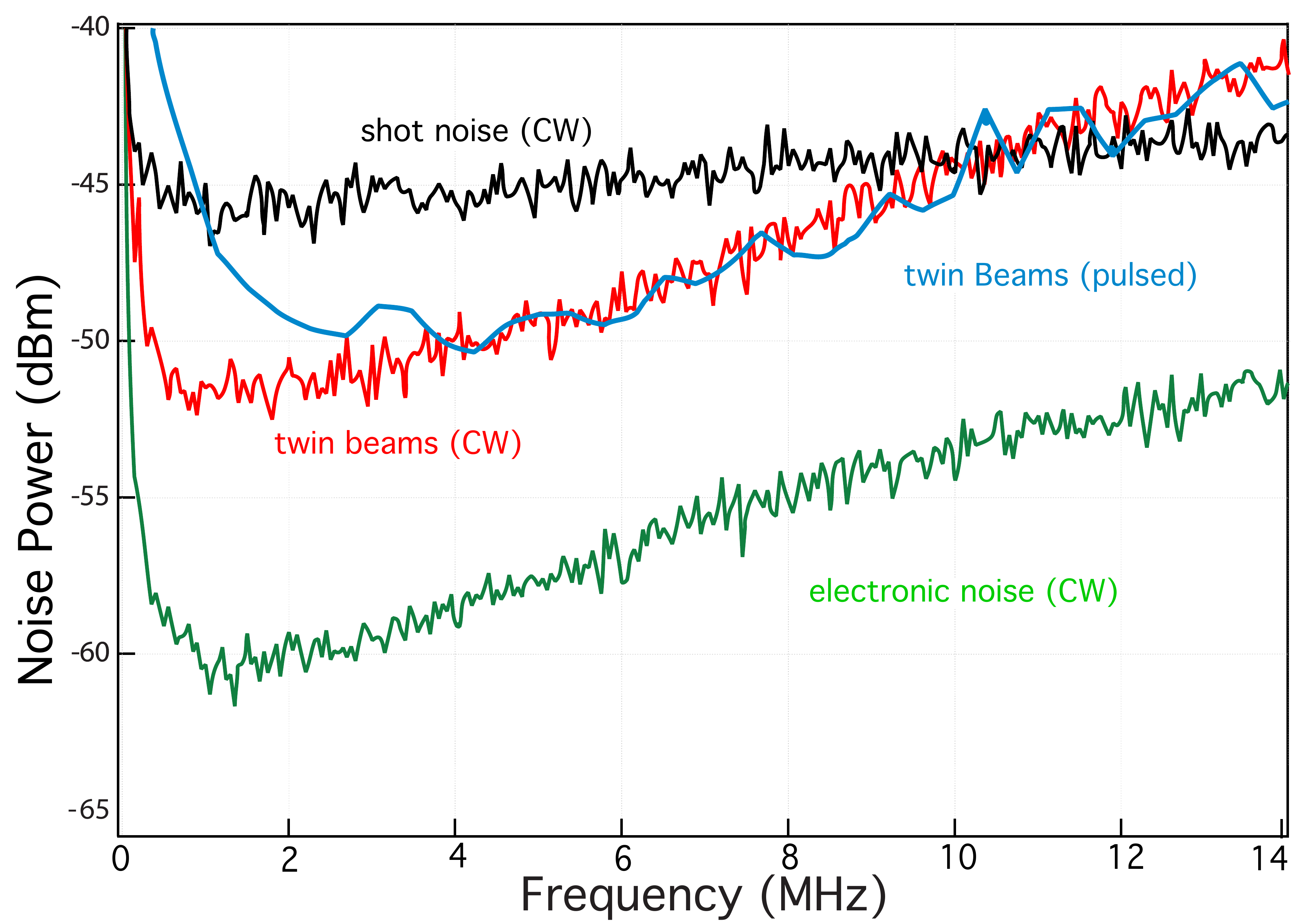}
\caption{\label{fig:pulseddataBB}Intensity-difference squeezing obtained for the the continuous-wave (red) and pulsed (blue) regimes. The shot noise (black) and electronic noise (green) are plotted for reference.
The data presented here are raw data without any corrections.
When these data are corrected for the electronic noise, the value of the shot noise is constant from 100 kHz to 15 MHz.}
\end{figure}
The average pulsed intensity-difference squeezing spectrum is plotted alongside CW measurements in Fig.~\ref{fig:pulseddataBB}.
The CW measurements for the shot noise and squeezing were taken without the high-pass filters pictured in Fig.~\ref{fig:setupBB} since the ringing effect discussed above does not apply to the CW case.
As shown in Fig.~\ref{fig:pulseddataBB}, the intensity-difference noise power in the pulsed and CW regimes is similar above 3~MHz.
At lower analysis frequencies, however, we observe a discrepancy between these two regimes in the form of a large excess noise for the pulsed measurements.
We suggest that the excess noise on the squeezing partially comes from the fact that the residual ringing at the detection stage described earlier is not completely filtered by the high-pass filter. 
At frequencies larger than 10~MHz, CW and pulsed squeezing vanishes and the noise level rises above SNL.
This effect is due to a different propagation velocity of the probe and the conjugate inside the atomic medium and can be easily mitigated using an electronic delay line\cite{Machida89}.

\section{Vacuum-squeezed twin beams}
\begin{figure}[htbp]
\centering\includegraphics[width=0.85\columnwidth]{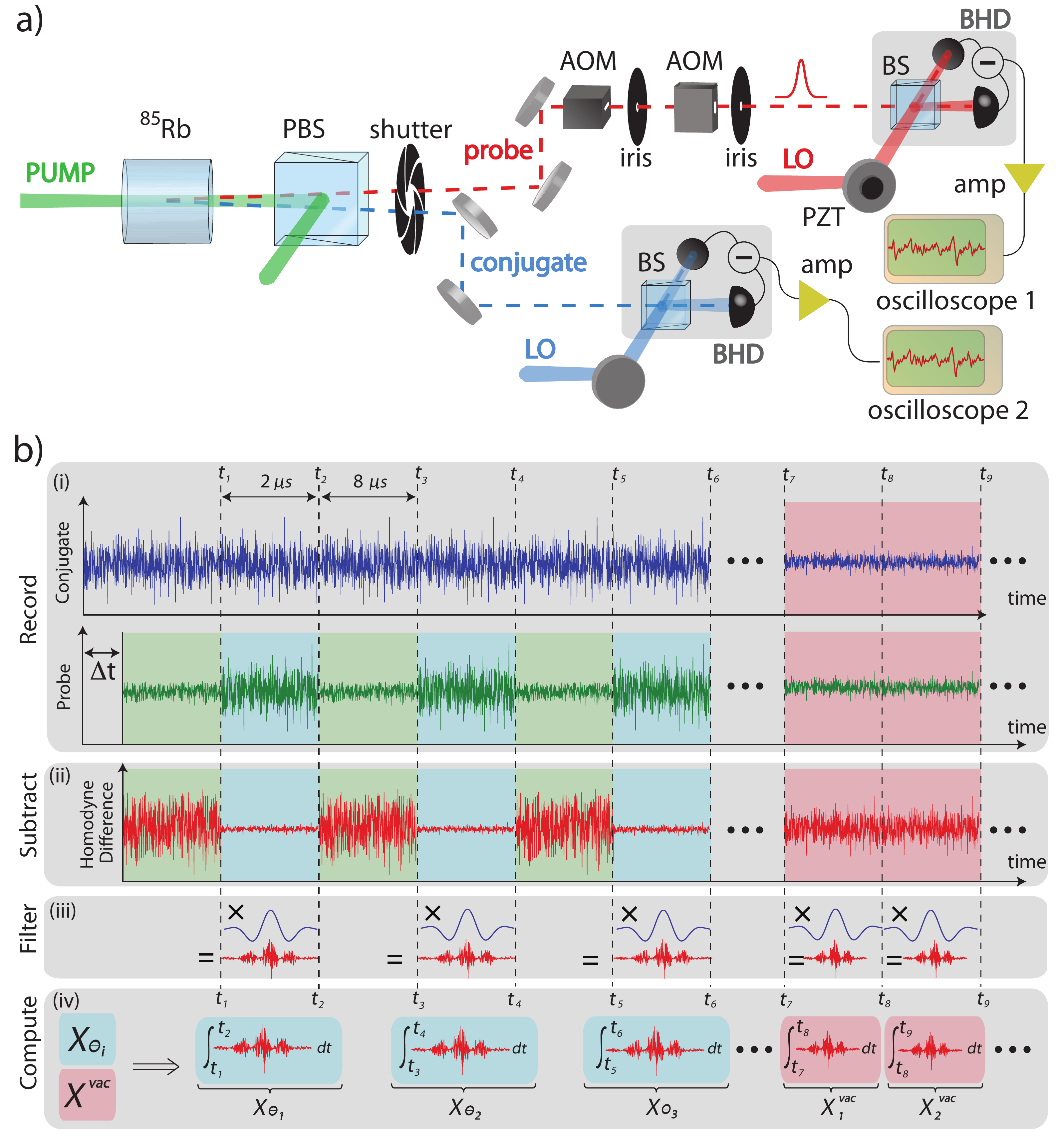}
\caption{\label{fig:setupVS} (a) Detection of vacuum-squeezed twin beams.
(b) Procedure to measure joint quadratures.  
(i) The CW homodyne detection of the conjugate (blue trace) samples the conjugate generalized quadrature $X^c_{\theta^c}$. 
 The probe homodyne detection (green trace) alternates between detecting the probe generalized quadrature $X^p_{\theta^p}$ (blue regions) and vacuum (green regions).
Probe and conjugate traces are recorded.
An offset time $\Delta t$ is introduced when the data are analyzed in order to maximize the squeezing.
 (ii) The difference signal (red trace) between the probe and conjugate homodyne detection alternates between the joint quadrature $X^{-}_{\theta}$ (blue regions) and the sum of the conjugate excess noise and the vacuum noise (green regions).
Similar operations are needed to measure $X^{+}_{\theta}$.
 (iii) We spectrally filter $X^{-}_{\theta}$ by multiplying the blue region with the window function of Eq.~\ref{eq:window}.  
(iv)  The values of $X^{-}_{\theta}$ are obtained after integration over the blue regions.  
At the end of the measurement, the shutter controlling the probe and conjugate fields is closed and the shot noise (red regions) is measured analogously. Traces in these figures are not experimental results and are included only to clarify the analysis procedure. }
\end{figure}

In order to demonstrate that the beams are entangled, we would like to generate pulsed vacuum-squeezed twin beams suitable for homodyne detection.
To obtain these pulsed beams, we first generate CW vacuum-squeezed twin fields (Fig.~\ref{fig:4WM}b) and direct the probe field through two AOMs that act as a switch.
These AOMs are driven with synchronously pulsed RF power to deflect the mode when the squeezed vacuum is to be turned off, and oriented such that the acoustic waves  propagate at right angles to each other to increase the extinction of the mode(Fig.~\ref{fig:setupVS}a).
The conjugate beam remains CW.
The probe and conjugate are then detected with separate homodyne detectors with CW LOs and the quadrature values recorded.
It should be noted that pulsing the 4WM pump beam is not an option to produce pulsed vacuum-squeezed twin beams since the 4WM process typically requires several $\mu$s of pumping time to reach steady state, which is on the time scale of interest for the pulses.

Nonzero diffracted orders emerging from the AOMs are blocked with irises placed immediately after each AOM.
We choose to detect the non-diffracted order since it suffers less loss and less mode distortion than the deflected beams.
Preserving the spatial mode quality of the probe is critical since any mode mismatch between the probe field and the probe LO used in the homodyne detection contributes to an additional effective loss.
The disadvantage of this technique is that extinction of the probe is limited by the diffraction efficiency that can be reached by the AOMs.
Orienting the AOMs perpendicularly to one another allowed us to achieve 97\% suppression of the probe mode, when activated, and a transmission greater than 95\%, when not activated.

To measure the noise of the phase-dependent joint quadratures $X^{-}_{\theta}$ and $X^{+}_{\theta}$, we pulse the probe field $10^4$ times over 100 ms with a period of 10~$\mu$s (Fig.~\ref{fig:setupVS}b) as we sweep the phase of the probe LO by $\pi$.
This technique assumes a certain degree of phase stability of the probe LO over the measurement time \cite{Appel:2008vl,Honda:2008cj,Lvovsky:2009fr}.
To verify that the phase excursion between the probe and conjugate  is negligible over the relevant timescale, we seed the twin beams with a coherent state and detect the resulting classical interference with the LOs.
Over periods of 100 ms, we typically obtain a phase standard deviation on the order of 1 degree.

When the probe is turned on, the homodyne difference signal yields a detection of $X^{-}_{\theta}$ and the sum of $X^{+}_{\theta}$.
When the probe is off, the probe homodyne detector measures shot noise and the conjugate homodyne detector measures the uncorrelated generalized quadrature $X^{c}_{\theta^{c}}$.
The difference and sum of these two contributions yields the same noise power in excess of the two-beam SNL.
At the end of the 100 ms sequence, the twin fields are blocked with a mechanical shutter and we record 10 ms of two-beam vacuum shot noise.

As discussed elsewhere \cite{McCormick:2007tl,Glorieux:2010ja}, the 4WM process exhibits a frequency-dependent squeezing spectrum over a bandwidth of approximately 20 MHz.
In order to study the phase-dependence of the noise power of the joint quadratures, we wish to confine our attention to a limited bandwidth where the squeezing is maximized.
To achieve this goal, we multiply each portion of the time trace corresponding to the measurement of the joint quadratures (blue regions in Fig.~\ref{fig:setupVS}b-iii) by the window function : 
\begin{equation}
\label{eq:window}
W(t)=\frac{1}{\sigma \sqrt{2\pi}}  \cos(\omega_{0} (t-t_{0}))  e^{\frac{-(t-t_{0})^{2}}{2\sigma^{2}}}  \Theta(\tau/2-|t-t_0|)
\end{equation}
$\Theta(t)$ refers to the Heaviside step function, $\tau$ the probe pulse width, and $t_{0}$ the midpoint of the probe pulse in time.
According to the convolution theorem, multiplying the noise time trace by $W(t)$ allows us to study a band of frequencies in the squeezing spectrum centered around $\omega_{0}/2\pi$ with a $\frac{1}{e^{2}}$ spectral width of $\sigma^{-1}$.
CW measurements of the squeezing reveal that the squeezing is maximized over a bandwidth of approximately 300 kHz centered around 750 kHz.
Accordingly, we selected a value of 0.5~$\mu s$ for the window width $\sigma$. 
\begin{figure}[]
\centering\includegraphics[width=0.65\columnwidth]{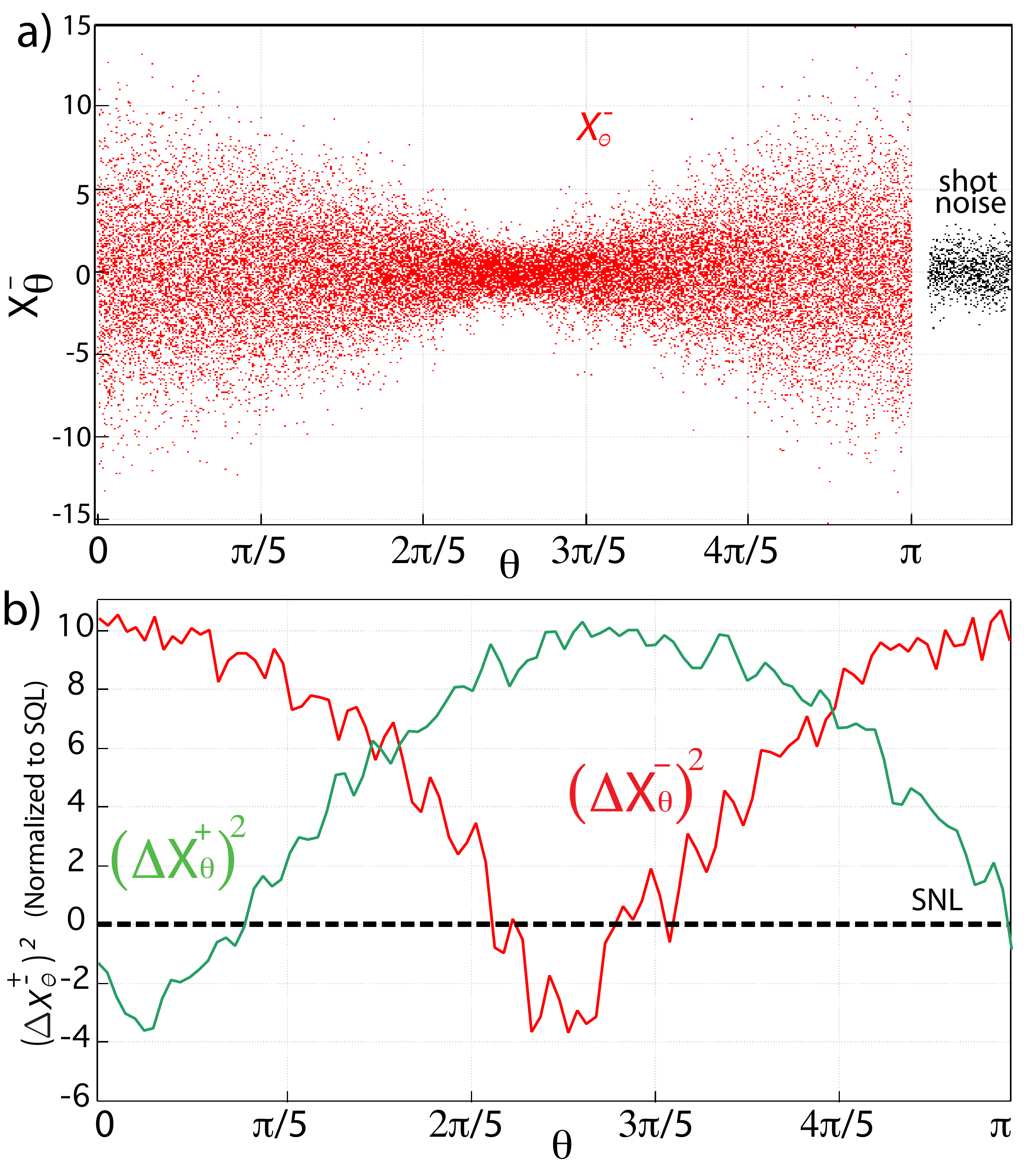}
\caption{\label{fig:pulseddataVS}Detected vacuum squeezing. (a) Normalized values for the joint quadrature $X^{-}_{\theta}$ of the twin beams (red) as a function of the joint quadrature phase $\theta$ and two-beam vacuum shot noise calibration (black). (b) The measured quadrature values are binned over 100 equally spaced phases of $\theta$, and the variance $(\Delta X^{-}_{\theta})^{2}$ of each bin is plotted against the 100 corresponding average values of $\theta$.  The black line indicates the shot noise limit (SNL).
}
\end{figure}
We plot $10^4$ recorded values of $X^{-}_{\theta}$ versus $\theta$ in Fig.~\ref{fig:pulseddataVS}a  (red points)alongside a sampling of the twin beam shot noise (black points).
The same data were recorded for  $X^{+}_{\theta}$ but are not shown for the sake of clarity.
To quantify the level of squeezing, the red points in Fig.~\ref{fig:pulseddataVS}a are binned over 100 equally-spaced phases of the generalized quadrature $\theta$.
The variance of the $X^{-}_{\theta}$ and $X^{+}_{\theta}$ values in each bin was computed and plotted against the average joint quadrature phase $\theta$ of each bin in Fig.~\ref{fig:pulseddataVS}b.
The SNL is obtained by computing the variance of the black points in Fig.~\ref{fig:pulseddataVS}a.
From the figure, we observe 3.8~$\pm$~0.2 dB of squeezing for both $X^{-}_{\theta^{-}}$ and $X^{+}_{\theta^{+}}$.
The uncertainty of 0.2~dB is obtained by calculating the standard deviation of the noise power over a phase range of $\pi/20$ around the phase $\theta=\pi/2$.
The measured inseparability is $I=0.83\pm 0.04<2$, which is sufficient to conclude that the beams are entangled.

A more stringent criterion is commonly used to describe a higher degree of entanglement, known as EPR entanglement.
This criterion quantifies the knowledge about the state of the probe beam by measuring the conjugate.
Although conditional variances are needed to quantify EPR entanglement, for Gaussian twin beams it has been shown that the product of the conditional variances is bounded by $4\left<(\Delta X^{-}_{\theta^{-}})^2\right>\left<(\Delta X^{+}_{\theta^{+}})^2\right>$ and therefore a value smaller than 1 for this product establishes EPR entanglement.
From the data shown on Fig.~\ref{fig:pulseddataVS}b, we obtain an upper bound for the product of conditional variances of  $0.69\pm 0.07<1$, which is sufficient to verify the presence of EPR levels of entanglement.\\

\section{Conclusion}
In this paper we demonstrate a source of pulsed bipartite entangled beams based on four-wave-mixing in a hot atomic vapor.
Substantial improvement in the degree of correlation ($3.8\pm 0.2$~dB below the SNL) and bandwidth (MHz) compared to previous work with an atomic vapor is reported.
Inseparability and EPR levels of entanglement are demonstrated with this technique in the pulsed regime.
While comparable results have been reported using the Kerr effect in an optical fiber at 1.5$\mu$m \cite{PhysRevLett.86.4267}, in this paper we demonstrate pulsed beams that match the bandwidth and the wavelength of atomic quantum memory techniques involving Rb vapor.
Finally, as 4WM is known to exhibit multi--spatial--mode properties \cite{Boyer:2008ts}, we think that this source will be of great interest for spatially multimode atomic memories \cite{Glorieux:2012uo}.

\section*{Acknowledgements}
This work is supported by the AFSOR and the Physics Frontier Center at the NIST/UMD Joint Quantum Institute.
\section*{References}
\bibliography{biblio}
\end{document}